\documentclass{article}

\usepackage{PRIMEarxiv}
\usepackage{amssymb}
\usepackage{amsmath}
\usepackage{longtable}
\usepackage{lipsum} 
\usepackage{mathptmx}
\usepackage{textcomp}
\usepackage{algorithm}
\usepackage{mdwmath}
\usepackage{mdwtab}
\usepackage{array}
\usepackage{graphics}
\usepackage{algorithmic}
\usepackage{eqparbox}
\usepackage{filecontents}
\usepackage{hhline}
\usepackage{multirow}
\usepackage{balance}
\usepackage{lscape}
\usepackage{lineno}
\usepackage{placeins}
\usepackage{float}
\usepackage{changes}
\usepackage{csquotes}
\usepackage{longtable}
\usepackage{lipsum} 
\usepackage{textcomp}
\usepackage{algorithm}
\usepackage{mdwmath}
\usepackage{mdwtab}
\usepackage{array}
\usepackage{rotating}
\usepackage{graphics}
\usepackage{algorithmic}
\usepackage{eqparbox}
\usepackage{empheq}
\usepackage{hhline}
\usepackage{multirow}
\usepackage{subcaption}
\usepackage{xcolor}
\usepackage{balance}
\usepackage{lineno}
\usepackage{placeins}
\usepackage{float}
\usepackage{csquotes}
\usepackage{lscape}
\usepackage{url}            
\usepackage{booktabs}       
\usepackage{amsfonts}       
\usepackage{nicefrac}       
\usepackage{microtype}      
\usepackage{lipsum}
\usepackage{fancyhdr}       
\usepackage{graphicx}       
\graphicspath{{media/}}     

\title{An Explainable Attention Model for Cervical Precancer Risk Classification using
	Colposcopic Images
\thanks{\textit{\underline{Citation}}: 
\textbf{Khare et al. An Explainable Attention Model for Cervical Precancer Risk Classification using
	Colposcopic Images. Pages.... DOI:000000/11111.}} 
}

\author{  Smith K. Khare \\
  $^1$Applied AI and Data Science Unit, Mærsk Mc-Kinney Møller Institute\\
  Faculty of Engineering  \\
  University of Southern Denmark \\
  Odense, Denmark\\
    $^2$Centre for Clinical Artificial Intelligence\\
  Odense University Hospital  \\
  Odense, Denmark\\
  \texttt{smkh@mmmi.sdu.dk} \\
   \And
 Berit Bargum Booth \\
  Centre for Department of Gynecology and Obstetrics\\
Odense University Hospital \\
  Odense, Denmark\\
  \And
  Victoria Blanes-Vidal \\
  Applied AI and Data Science Unit, Mærsk Mc-Kinney Møller Institute\\
  Faculty of Engineering  \\
  University
of Southern Denmark \\
  Odense, Denmark\\
 \And
  Lone Kjeld Petersen \\
  Centre for Research Unit for Gynecology and Obstetrics (Odense)\\
  Odense University Hospital \\
  Odense, Denmark\\
\And
 Esmaeil S. Nadimi \\
  $^1$Applied AI and Data Science Unit, Mærsk Mc-Kinney Møller Institute\\
  Faculty of Engineering  \\
  University
of Southern Denmark \\
  Odense, Denmark\\
$^2$Centre for Clinical Artificial Intelligence\\
  Odense University Hospital  \\
  Odense, Denmark\\
  \texttt{smkh@mmmi.sdu.dk} \\}

\begin{document}
\maketitle

\begin{abstract}
Cervical cancer remains a major worldwide health issue, with early identification and risk assessment playing critical roles in effective preventive interventions. This paper presents the Cervix-AID-Net model for cervical precancer risk classification. The study designs and evaluates the proposed Cervix-AID-Net model based on patients colposcopy images. The model comprises a Convolutional Block Attention Module (CBAM) and convolutional layers that extract interpretable and representative features of colposcopic images to distinguish high-risk and low-risk cervical precancer. In addition, the proposed Cervix-AID-Net model integrates four explainable techniques, namely gradient class activation maps, Local Interpretable Model-agnostic Explanations, CartoonX, and pixel rate distortion explanation based on output feature maps and input features. The evaluation using holdout and ten-fold cross-validation techniques yielded a classification accuracy of 99.33\% and 99.81\%. The analysis revealed that CartoonX provides meticulous explanations for the decision of the Cervix-AID-Net model due to its ability to provide the relevant piece-wise smooth part of the image. The effect of Gaussian noise and blur on the input shows that the performance remains unchanged up to Gaussian noise of 3\% and blur of 10\%, while the performance reduces thereafter. A comparison study of the proposed model's performance compared to other deep learning approaches highlights the Cervix-AID-Net model's potential as a supplemental tool for increasing the effectiveness of cervical precancer risk assessment. The proposed method, which incorporates the CBAM and explainable artificial integration, has the potential to influence cervical cancer prevention and early detection, improving patient outcomes and lowering the worldwide burden of this preventable disease.
\end{abstract}


\section{Introduction}\label{sec1}
Cervical cancer is the fourth leading cause of death among female malignancies, with high morbidity and mortality rates if diagnosed in late stages, mainly affecting sexually active adult women aged over 30 years \cite{INT4, INT5, INT6}. The cervix is part of the female reproductive system, located in the lowest fibromuscular section of the uterus and accessible for inspection and sampling though the vagina. The cervix has different linings. The endocervical canal is lined with glandular epithelium, and the ectocervix is lined with squamous epithelium. The squamous epithelium meets the glandular epithelium at the squamocolumnar junction (SCJ).
The SCJ moves during early adolescence and during a first pregnancy. The original SCJ originates in the endocervical canal, but over time, the SCJ comes to lie on the ectocervix and becomes the new SCJ \cite{fadare2019normal}. In colposcopy terminology, the SCJ is this new SCJ. The epithelium between these two SCJs is the transition zone (or transformation zone, TZ), and its position is variable, depending on factors such as age, hormonal status, birth trauma, use of oral contraceptives, and pregnancy  \cite{INT1, Prendiville2017}. Colposcopically the TZ is classified as: Type 1, Type 2 and Type 3, depending on its visibility \cite{INT2, INT3}. Cervical cancer is preceded by Cervical Intraepithelial Neoplasia (CIN) which arises in the TZ. In the year 2018, the World Health Organization (WHO) issued a worldwide call to eliminate cervical cancer \cite{INT7}. Cervical cancer develops from CIN lesions over years making time for screening, diagnosis and preventive treatment. There is an urgent need for accurate and timely detection of cervical cancer. This is particularly so in low- and middle-income nations, where severe poverty and gender discrimination significantly restrict a woman’s ability to seek care, accounting for almost 88\% of cervical cancer fatalities \cite{INT8}.

Many screening methods are available today to detect precancerous lesions and abnormal growth of epithelial cells on the cervix. These methods include primary screening with cervical cytology (Pap tests), the human papillomavirus (HPV) test or co-testing. Cervical cytology requires competent cytologists to perform microscopic analyses, often unavailable in low-resource settings \cite{INT9}. The HPV test is recommended by the WHO strategy due to the high sensitivity of the test, but the associated lower specificity increases the number of screen positive females referred for secondary screening by colposcopy and adds to the burden of specialists. Medical authorities recommend colposcopy as the gold standard for assessing cervical cancer precursors. Examining colposcopy images is time-consuming and requires trained medical specialists. Even skilled colposcopist miss cervical precancerous lesions in need for preventive treatment in 40\% of examinations \cite{https://doi.org/10.1002/ijc.25470}.  In recent years, automated decision-making using medical imaging techniques has increased multi-fold due to the advancement of artificial intelligence (AI). Automated decision-making AI models has reflected an immense potential for the detection of malignant tumors  \cite{LS1, LS2, LS3}. These advancements in AI have also attracted colposcopists to get assistance from AI in clinical decision-making. Therefore, there is an increasing interest in using AI to automate colposcopy image assessments, so that, identifying the presence of precancerous or cancerous cells in the cervix, becomes feasible at a large scale, with higher speed and lower costs, than nowadays. 
\section{Literature review}\label{sec2}
The literature review discusses the recently developed automated decision-making AI models for assessment of cervical precancer. Li et al. \cite{1} implemented a graph convolutional network with edge features (E-GCN) for classifying negative (Neg) and positive (Pos) classes. The positive class is comprised of low-grade squamous intraepithelial lesions or worse (LSIL+), while the negative class includes non-cancerous cases. Elakkiya et al. \cite{2} presented a hybrid model called faster small-object detection neural networks (FSOD-GAN)  combining faster region-based convolutional neural network (FR-CNN) and generative adversarial network (GAN) to detect normal cervical images (categorized as type 1, type 2 and type 3), and abnormal (AN) cervical images categorized based on cervical cancer severity, stage-I, stage-II, and stage-III.  Adweb et al. \cite{3} utilized pre-existing 18-layer residual networks to classify healthy control (HC) and cervical images from  pre-cancerous lesions (PC). The authors tested their model by combining 4000 PC images from one dataset and 800 from another, translated (augmented) to 1920 images. Kim et al. \cite{5}  developed a convolutional neural network (CNN)-based architecture called  AIDOT to classify normal, CIN1, CIN2/CIN3, and cancer classes. Saini et al. \cite{6} developed a CNN-based model called  ColpoNet, which has been motivated by the DenseNet model. The authors performed binary classification of colposcopic images in Type 1, which consists of normal and CIN1, while Type 2 includes CIN2 or worse, respectively. Habtemariam et al. \cite{7} used an existing EfficientNetB0-based CNN model to perform cervix Type 1 TZ, Type 2 TZ, and Type 3 TZ classification. Ma et al. \cite{8} performed four class classifications to detect normal, LSIL, high-grade squamous intraepithelial lesion (HSIL), and cancer. The authors combined CNN-based segmentation, following color and Haralick texture feature extraction, and a multi-modal classifier model. Binhua et al. \cite{9} developed colposcopy-based classification and diagnosis of cervical lesions using a dense U-NET model to detect CIN1, CIN2, and CIN3. Kim et al. \cite{11} developed a binary classification model to classify normal/CIN1 and CIN2 or  worse (CIN2+) using cervigram images (now discontinued). Cho et al. \cite{12} used Inception-Resnet-v2 to classify high-risk (CIN2, CIN3, and cancer) and low-risk (NC and CIN1) CINs; while Resnet-152 to classify the low-risk squamous intraepithelial lesions (LSIL and NC) and high-risk squamous intraepithelial lesions (cancer and HSIL). Liu et al. \cite{13} developed a two binary classification model for classifying NC versus LSIL+ and HSIL- versus HSIL+ using the residual network-50 model. Chandran et al. \cite{14} developed a colposcopy ensemble network (CYENET) to detect Type 1 TZ, Type 2 TZ, and Type 3 TZ cervical cancer in the colposcopy images. Wu et al.  \cite{15} developed a colposcopic artificial intelligence auxiliary diagnostic system (CAIADS) using the CNN model to detect CIN2+, CIN3+, and cancer. Sim{\~o}es et al. \cite{17} used a hybrid neural network based on Kohonen's self-organizing maps and multi-layer perceptron (MLP) model to classify dot patterns in colposcopic images. Asiedu et al. \cite{18} developed an automated model using Gabor segmentation followed by Haralick’s texture, color space transformation calculations, and lesion size estimation features. A support vector machine classifier was used to classify these features into positive class (CIN1, CIN2, CIN3, and invasive cancer) and negative class (normal, condiloma, and cervicitis). Miyagi et al. \cite{19} proposed the classification of uterine cervical squamous epithelial lesions from colposcopy images. Their method used a CNN model to classify low-grade (consisting of CIN1) and high-grade (combining CIN2 and CIN3) classes.  Song et al. \cite{20} proposed a binary classification model for classifying negative class (CIN1) and positive class (CIN2/3+). The authors used a multi-modal entity coreference CNN model to detect the corresponding classes using cervicography images. Miyagi et al. \cite{21} used the CNN model for the binary classification of CIN1 and CIN2+ using colposcopic images.

The literature review reveals that the majority of existing methods used pre-existing models for the detection of cervical lesions. Besides, most of existing automated decision-making models have limited performance concerning accuracy and other evaluation matrices (refer Table~\ref{tab6}). Also, most studies validate their model's performance using holdout (HO) validations, which are highly prone to over-fitting and biases. Additionally, only a few studies have explored explainable artificial intelligence (XAI), which attempts to explain model decisions. This demands a need for an accurate and explainable model for cervical precancer risk classification. Therefore, this study presents a novel and explainable cervical precancer risk classification model using colposcopic images. The model uses a combination of five convolutional layers and five Convolutional Block Attention Module (CBAM) to develop the proposed Cervix-AID-Net model. Also, the proposed Cervix-AID-Net model integrates gradient class activation maps (Grad-CAM), Local Interpretable Model-agnostic Explanations (LIME), CartoonX, and pixel rate distortion explanation (RDE). The significant contributions of the paper are listed as follows:

\begin{itemize}
	\item We propose a simple, lightweight, and effective cervical precancer risk classification model using CBAM that can be widely applied to boost the representation power of CNNs.
	
	\item To the best of our knowledge, we are the first group to present the integration of Grad-CAM, LIME,  CartoonX, and pixel RDE for the explanations.
	\item In this paper we evaluate the amount of distortion required for each module to flip the decision from one class to another. 
\end{itemize}
The overall paper is organized as, Section~\ref{sec1} presents an introduction, literature review in Section~\ref{sec2}, details about the Cervix-AID-Net model in Section~\ref{sec3}, experimental setup and results in Section~\ref{sec4}, Discussion in Section~\ref{sec5}, and finally, Section~\ref{sec6} concludes the paper. 
\section{Methodology}
\begin{figure}[!htbp]
	\centering
	\includegraphics[width=0.8\textwidth]{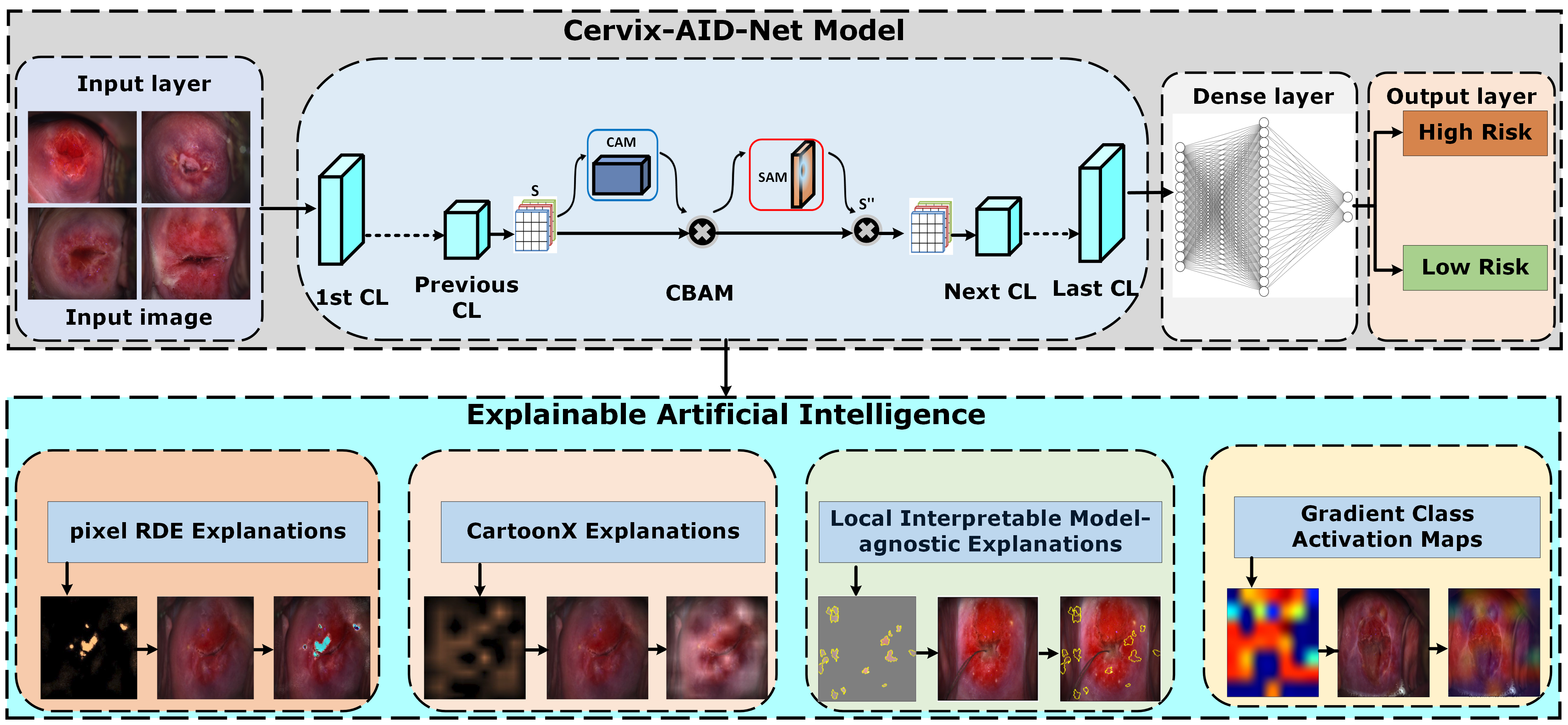}
	\caption{The proposed explainable Cervix-AID-Net model for high-risk and low-risk classification.}
	\label{fig1}
\end{figure}

The description of the proposed model comprises four subsections: dataset description, details of the proposed Cervix-AID-Net model, performance evaluation, and explainable AI. Figure~\ref{fig1} shows an overview of the proposed Cervix-AID-Net. 
\subsection{Dataset details}
All data and images were collected from patients examined using a DYSIS colposcope Version 3 at the Department of Gynecology and Obstetrics, Randers Regional Hospital, Denmark between 2017-2020. Colposcopy examinations were included if the colposcopists had reported a full or partially visible transformation zone, and four cervical biopsies taken. Images not suitable for annotation were excluded (blurry, too much light, mucous or blood covering visible changes). Based on international guidelines for risk assessment and treatment of cervical cancer precursors histological diagnoses of normal, inflammation and CIN1 were considered to represent low-grade disease. Other histological diagnoses present were CIN grade 2, CIN3, carcinoma in situ (CIS), adenocarcinoma in situ (AIS) and squamous cell carcinoma (SCC) were classified to represent high-grade disease. Images from women with a diagnosis of only ungradable CIN were removed (n=1). Women who had one or more other diagnosis were still included and the worst of the other diagnoses was considered to be the grade of dysplasia present.  \cite{perkins20242019}. The details of the dataset is shown in Table~\ref{dataset}. 

\begin{table}  \linespread{1.2}\selectfont\centering
	\centering
	\caption{Details of the dataset used to test the proposed model.}  \label{dataset}   
	\begin{tabular}   { p{2.6cm} | p{3.3cm} |c}
		\hline\hline 	
		& & Number N(\%) \\ \hline
		Total (X+Y)	&		&	178	\\ \hline
		Age median (range)	&		&	30.4 (20.0-62.7)	\\ \hline
		BMI median (range)	&		&	22.8 (18.4-47.6)	\\ \hline
		\multirow{4}{*}{Smoking}	&	No	&	96 (53.9\%)	\\
		&	Current	&	41 (23\%)	\\
		&	Previous	&	40 (22.5\%)	\\
		&	Unknown	&	1 (0.6\%)	\\ \hline
		\multirow{5}{*}{Contraception use}	&	Oral	&	77 (43.2\%)	\\
		&	IUD	&	28 (15.7\%)	\\
		&	Condom	&	9 (5.1\%)	\\
		&	Other	&	3 (1.7\%)	\\
		&	None	&	61 (34.3\%)	\\ \hline
		\multirow{4}{*}{HPV vaccination}		&	Not vaccinated	&	61 (34.3\%)	\\
		&	Vaccinated	&	108 (60.7\%)	\\
		&	On-going	&	8 (4.5\%)	\\
		&	Unknown	&	1 (0.5\%)	\\ \hline
		\multirow{4}{2.6cm}{New Referral (X=137)}	&	Low-grade (ASCUS/LSIL)	&	80 (58.4\%)	\\
		&	High-grade (ASC-H/AGC/HSIL)	&	57 (41.6\%)	\\ \hline
		\multirow{4}{2.6cm}{HPV test with referral (From 137)}	&	Negative	&	2 (6.3\%)	\\
		&	HPV 16 + other hr HPV	&	3 (9.4\%)	\\
		&	Other hr HPV alone	&	26 (81.2\%)	\\
		&	HPV 18 alone	&	1 (3.1\%)	\\ \hline
		HPV status unknown & & 146 (82.02\%) \\ \hline
		\multirow{4}{2.6cm}{Follow-up due to (Y=41)}	&	CIN 1	&	3 (7.3\%)	\\
		&	CIN 2	&	31 (75.6\%)	\\
		&	Ungradable CIN	&	6 (14.6\%)	\\
		&	Unknown	&	1 (2.4\%)	\\ \hline  \hline
		
	\end{tabular} 
	\vspace{1mm}
	
	\tiny{BMI: Body Mass Index, IUD: intra-uterine device; ASCUS: atypical squamous cells of undetermined significance; ASC-H: atypical squamous cells favoring high grade; AGC: atypical glandular cells}
\end{table}

\subsection{Cervix-AID-Net}\label{sec3}
Attention has a significant influence on human perception \cite{730558}. The key feature of human visual systems is that they do not process entire scenes at once. Instead, humans use a series of partial glimpses, deliberately focusing on crucial regions, to better capture visual structure \cite{NIPS2010}. Inspired by this, we developed a lightweight CNN-based CBAM module to classify high-risk and low-risk cervical precancer. 
\subsubsection{Convolutional Block Attention Module (CBAM)}
\begin{figure}[!htbp]
	\centering
	\includegraphics[width=0.8\textwidth]{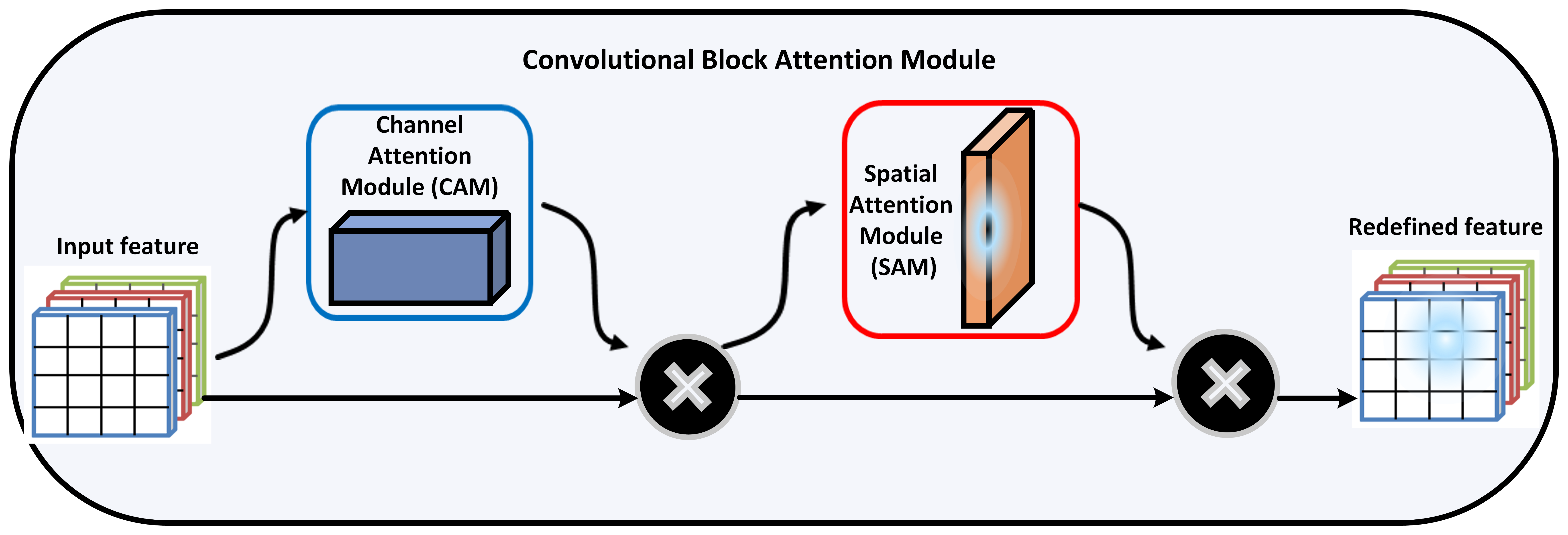}
	\caption{Schematic of convolutional block attention module (CBAM).}
	\label{pipeline}
\end{figure}

Fig.~\ref{pipeline} illustrates the overview of the CBAM. It comprises two successive sub-modules: channel-attention and spatial-attention \cite{CBAM}. The CBAM adapts to enhance the intermediate feature map at each convolutional block of deep networks.
The CBAM consecutively evaluates 1D channel attention map ${\textbf{M\textsubscript{ch}}} \in \mathbb{R}^{C\times 1\times 1}$ and a 2D spatial attention maps ${\textbf{M\textsubscript{sp}}} \in \mathbb{R}^{1\times H\times W}$ for a given intermediate feature map ${\textbf{S}} \in \mathbb{R}^{C\times H\times W}$ as input. The complete attention process can be described as: 
\begin{equation}
	\begin{split}
		\textbf{S}^{\prime} = \textbf{M\textsubscript{ch}(S)} \odot \textbf{S} \\
		\textbf{S}^{\prime\prime} = \textbf{M\textsubscript{sp}}(\textbf{S}^{\prime}) \odot \textbf{S}^{\prime}
	\end{split}
\end{equation}
where \textbf{M\textsubscript{sp}} and \textbf{M\textsubscript{ch}} are spatial and channel attention map, $C, H, \text{and}\,\, W$ are the channel, height, and width of input feature map, and $\odot$ is element-wise multiplication (Hadamard product). 
\subsubsection{Channel attention module (CAM)}
\begin{figure}[!htbp]
	\centering
	\includegraphics[width=0.8\textwidth]{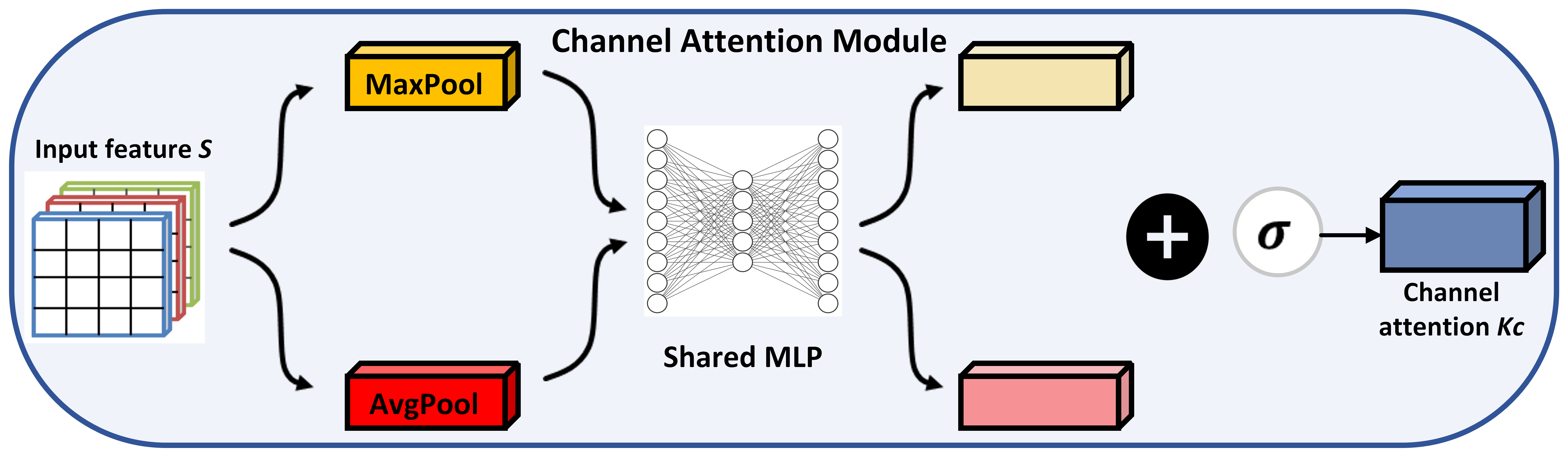}
	\caption{Schematic of channel attention module (CAM).}
	\label{cam}
\end{figure}

Each channel in a feature map serves as a feature detector, with channel attention focusing on `what' is significant to an input image. Fig.~\ref{cam} shows a graphical illustration of the steps involved in computing CAM. As shown in Fig~\ref{cam}, the spatial size of the input feature is squeezed to optimize the channel attention. Therefore, CAM uses average-pooled and max-pooled features simultaneously. The CAM module is evaluated as \cite{CBAM}:
\begin{equation}
	\begin{split}
		\textbf{M}_{\textbf{ch}}(\textbf{S}) = sigmoid(MLP(AVP(\textbf{S}))+MLP(MP(\textbf{S}))) \\
		= sigmoid(\textbf{W}_1(\textbf{W}_0(\textbf{S}_{\textbf{avg}}^{\textbf{ch}}))+\textbf{W}_1(\textbf{W}_0(\textbf{S}_{\textbf{max}}^{\textbf{ch}})))
	\end{split}
\end{equation}
where \textbf{W\textsubscript{0}} and \textbf{W\textsubscript{1}} are the weights of \textit{MLP}, spatial context descriptors generated by average-pooling and max-pooling are denoted by $\textbf{S}_{\textbf{avg}}^{\textbf{ch}}$ and $\textbf{S}_{\textbf{max}}^{\textbf{ch}}$, $\textbf{M}_{\textbf{ch}}$ is channel attention map, $AVP$ is average-pooling, $MP$ is max-pooling, and $MLP$ denotes a shared multi-layer perceptron network with one hidden layer. 
\subsubsection{Spatial attention module (SAM)}
\begin{figure}[!htbp]
	\centering
	\includegraphics[width=0.8\textwidth]{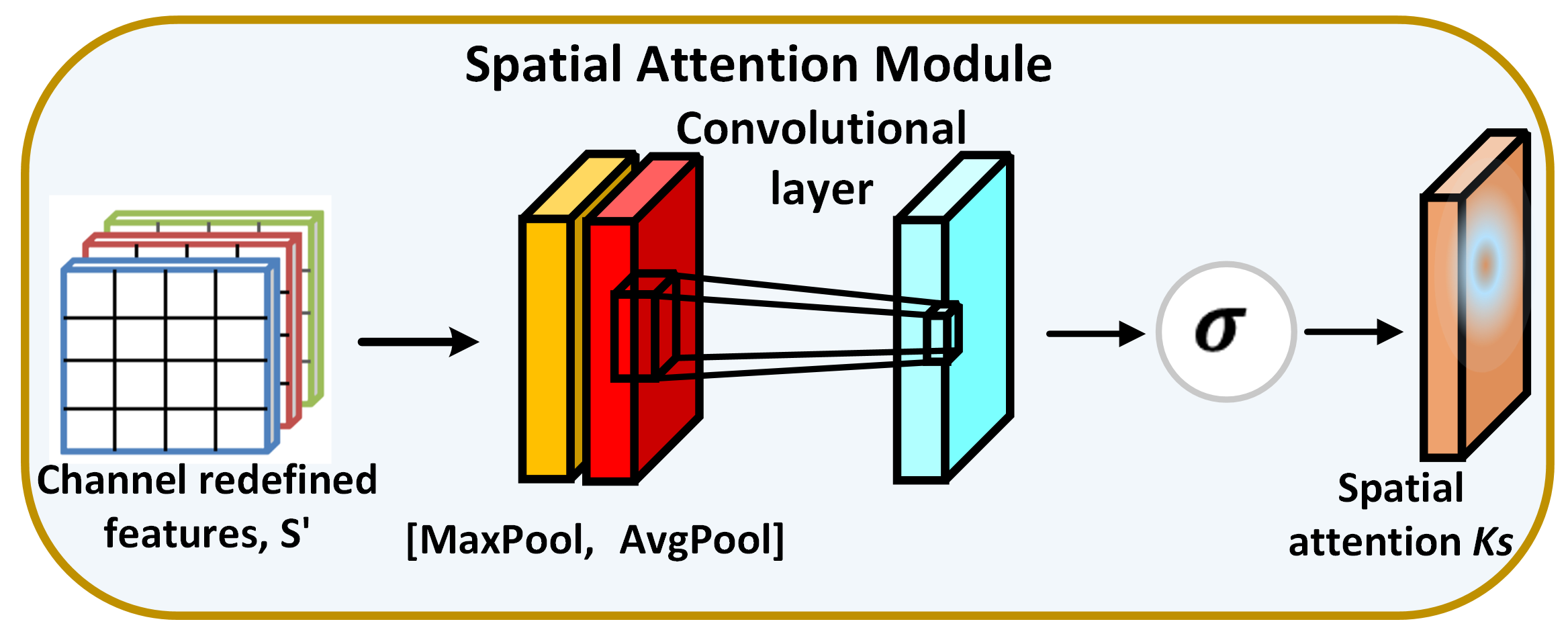}
	\caption{Schematic of spatial attention module (SAM).}
	\label{sam}
\end{figure}

CAM focuses on `what' is significant to an input image, whereas SAM focuses on `where' an informative part of an image is located. The evaluation of spatial attention covers applying average-pooling and max-pooling along the channel axis and concatenating them to extract representative features. After concatenation, a convolutional layer is applied to generate SAM, $\textbf{M}_{\textbf{sp}}(\textbf{S}) \in \mathbb{R}^{H \times W}$. Fig.~\ref{sam} shows the steps involved in computing the SAM. The mathematically spatial attention module can be evaluated as \cite{CBAM}:   
\begin{equation}
	\begin{split}
		\textbf{M}_{\textbf{sp}}(\textbf{S}) = sigmoid(f^{7\times 7}([AVP(\textbf{S});MP(\textbf{S})])) \\
		= sigmoid(f^{7\times 7}([\textbf{S}_{\textbf{avg}}^{\textbf{sp}};\textbf{S}_{\textbf{max}}^{\textbf{sp}}]))
	\end{split}
\end{equation}
where $\textbf{M}_{\textbf{sp}}$ is spatial attention map, $\textbf{S}_{\textbf{max}}^{\textbf{sp}}$ and $\textbf{S}_{\textbf{avg}}^{\textbf{sp}}$ are max-pooled and average-pooled feature maps across the channel, and $f^{7\times 7}$ represents convolutional operations using a filter size of 7. 
We developed a novel lightweight Cervix-AID-Net model using the CBAM block. Our Cervix-AID-Net model consists of five convolutional layers, five CBAM blocks, and three dense layers. Each convolutional layer is followed by a CBAM block to extract relevant feature maps. Table~\ref{tab3} represents the detailed Cervix-AID-Net model with its tuning and learning parameters. 

\begin{table}  \linespread{1.2}\selectfont\centering
	\centering \setlength\tabcolsep{3.5pt} 
	\caption{Tuning and learning parameters of the proposed Cervix-AID-Net model.}  \label{tab3}   
	\begin{tabular}   { c | c |c| c |c|c| c }
		\hline\hline 	
		Layer	&	Filters	&	KS	&	Padding	&	PS	&	ACT	&	PRM	\\ \hline \hline
		IL (Size:)	&	 224x224x3	&	--	&	--	&	--	&	--	&	--	\\ 
		CL	&	32	&	3x3	&	same	&	--	&	ReLU	&	896	\\ 
		CBAM	&	--	&	--	&	--	&	--	&	--	&	390	\\ 
		BN	&	--	&	--	&	--	&	--	&	--	&	128	\\ 
		MP	&	--	&	--	&	--	&	2x2	&	--	&	--	\\ 
		CL	&	64	&	3x3	&	same	&	--	&	ReLU	&	18496	\\ 
		CBAM	&	--	&	--	&	--	&	--	&	--	&	1194	\\ 
		BN	&	--	&	--	&	--	&	--	&	--	&	256	\\ 
		MP	&	--	&	--	&	--	&	2x2	&	--	&	--	\\ 
		CL	&	128	&	3x3	&	same	&	--	&	ReLU	&	73856	\\
		CBAM &	--	&	--	&	--	&	--	&	--	&	4338	\\ 
		BN	&	--	&	--	&	--	&	--	&	--	&	512	\\ 
		MP &	--	&	--	&	--	&	2x2	&	--	&	--	\\ 
		CL	&	384	&	3x3	&	same	&	--	&	ReLU	&	442752	\\ 
		CBAM	&	--	&	--	&	--	&	--	&	--	&	37394	\\ 
		BN	&	--	&	--	&	--	&	--	&	--	&	1536	\\ 
		MP	&	--	&	--	&	--	&	2x2	&	--	&	--	\\ 
		CL	&	256	&	3x3	&	same	&	--	&	ReLU	&	884992	\\ 
		CBAM	&	--	&	--	&	--	&	--	&	--	&	16770	\\ 
		BN	&	--	&	--	&	--	&	--	&	--	&	1024	\\ 
		MP	&	--	&	--	&	--	&	2x2	&	--	&	--	\\ 
		FL	&	--	&	--	&	--	&	--	&	--	&	--	\\ 
		DnL	&	Units: 256	&	--	&	--	&	--	&	ReLU	&	3211520	\\ 
		DnL	&	Units: 128	&	--	&	--	&	--	&	ReLU	&	32896	\\ 
		DpL	&	0.25	&	--	&	--	&	--	&	--	&	--	\\ 
		DnL	&	Units: 2	&	--	&	--	&	--	&	Softmax	&	258	\\ \hline
		&&&&&\textbf{Total}	&	\textbf{4729208}	\\  \hline \hline

	\end{tabular} 
	\vspace{1mm}
	
	\tiny{KS: Kernel size; PS: Pooling size; ACT: Activation; PRM: Parameters; ReLU: Rectified linear unit; IL: Input layer; CL: Convolutional layer; BN: Batch normalization; MP: Maxpooling layer, FL: Flatten layer; DnL: Dense layer; DpL: Dropout}
\end{table}
\subsection{Performance evaluation}
We used HO validation and ten-fold cross-validation (10-FCV) strategies to evaluate model performance. For HO validation, training uses 80\% data and the remaining for testing. Out  of 20\% testing data, 6\% for validation and 94\% for testing. The model training, validation, and testing utilized 3153 images. Therefore, the final train, validation, and test set contains 2524, 37, and 593 images, respectively. For 10F-CV, we randomly split the entire dataset into ten equal parts, of which training uses nine parts while the tenth part is for testing. The process is repeated ten times to estimate the average model performance. We have tested the Cervix-AID-Net model performance by evaluating accuracy (ACC), precision (PRC), negative predicted value (NPV), false positive rate (FPR), F1 measure (F1), sensitivity (SEN), specificity (SPF), balanced accuracy (B-ACC), and Mathew's correlation coefficient (MCC), details in \cite{canbek2022ptopi}.

\subsection{Explainable AI module}
Explainable AI (XAI) are AI systems and models that can offer transparent and intelligible explanations for their decision-making processes \cite{KHARE2024102019}. Traditional ML models, such as deep neural networks, are sometimes referred to as ``black boxes" since it can be hard to understand how they arrive at certain conclusions or predictions \cite{KHARE2023101898}. XAI facilitates transparency, interpretability, trust, and human-readable explanations in the decisions yielded by ML or DL models. We use four explainable modules: Grad-CAM, LIME, CartoonX, and pixel RDE to visualize and better understand the decisions of our proposed model. 
\subsubsection{Grad-CAM} The Grad-CAM technique operates on output feature maps. It identifies crucial regions in the input image by computing the gradients of the last convolutional layer, generating heat maps. These heat maps visually represent the significant regions in the input image, which helped the network to arrive at certain decisions \cite{gradCAM, ELDAHSHAN2024122388}. The mathematical formulation of grad-CAM technique is denoted by \cite{gradCAM}
\begin{equation}
	\beta^m_k = \frac{1}{Z} \Big(\sum_{i} \sum_{j} \frac{\partial y_m}{\partial F^{k}_{ij}}\Big)
\end{equation}
where $\frac{\partial y_m}{\partial F^{k}}$ represents the gradient score for class $m$ with respect to feature maps $F^k$. $\beta^m_k$ denotes  a partial linearization of the deep network downstream from $F$, and captures the `importance’ of feature map $k$ for a target class $m$. Finally, grad-CAM is obtained by taking a ReLU of a weighted combination of forward activation maps, denoted as \cite{gradCAM}:
\begin{equation}
	L_{Grad-CAM}^m =ReLU \Big(\sum_{k}\beta^m_kF^k\Big)
\end{equation}
\subsubsection{LIME} LIME is model-agnostic, which means it may be used on any machine/deep learning model independent of architecture or complexity \cite{KHARE2023110858}. The goal of LIME is to train surrogate models locally and explain a single prediction. It generates a synthetic dataset by randomly permuting samples from a normal distribution and collects predictions based on the opaque model to be explained. LIME uses the perturbed dataset to train an interpretable model. The mathematical formulation for LIME is denoted by \cite{LIME}
\begin{equation}
	\zeta(x) = \arg\min_{m \in \mathcal{M}} \mathcal{L}(g, m, \pi_x) + \Omega(m)
\end{equation}

where $\mathcal{M}$ is a class of Cervix-AID-Net, $\mathcal{L}$ is a fidelity function, and $\Omega(m)$ measures complexity of the explanations $m \in \mathcal{M}$.

\subsubsection{Pixel RDE} Pixel RDE are model-independent explanations inspired by rate distortion theory, which examines lossy data compression \cite{RDE}. In pixel RDE, explanations use a sparse mask to highlight relevant features from incoming data. The mask is tailored to minimize distortion in model output after perturbing unselected input features while remaining sparse. RDE tries to address the constrained optimization problem over a mask denoted by \cite{RDE}:
\begin{equation}
	\label {op: pixel rde} \min _{s_m\in \{0,1\}^n:\,\|s_m\|_0 \leq \ell } \; \mathop {\mathbb {E}}_{k\sim \mathcal {K}} \Big [ d\Big (\Psi (x), \Psi (x\odot s_m + (1-s_m)\odot k)\Big )\Big ] \end{equation}
where $\Psi : \mathbb{R}^n \rightarrow \mathbb{R}^m$ is a pre-trained Cervix-AID-Net model with $m (High-risk \, \text{and} \, Low-risk)$ classes, dimension of model input is represented by $n$, $x$ denotes relevant input features, sparse marks is represented by  $s_m\in \{0,1\}^n$,  $\mathcal{K}$ is a distribution  over input perturbations $k \in \mathbb{R}^n$, $\odot$ is element-wise multiplication (Hadamard product), $l \in \{1,2,...,n\}$ is a sparsity level for mask explanation $s_m$, and $d(\Psi (x))$ is a measure of distortion. However, in practice, RDE optimization problem is relaxed to continuous masks denoted by \cite{RDE}:

\begin{equation}
	\min _{s_m\in [0,1]^n} \mathop {\mathbb {E}}_{k\sim \mathcal {K}} \Big [ d\Big (\Psi (x), \Psi (x\odot s_m + (1-s_m)\odot k)\Big )\Big ] +\lambda  ||{s_m}||_1 
\end{equation}
where $\lambda$ takes values $> 0$, which is a hyper-parameter for the sparsity level.
\subsubsection{CartoonX} CartoonX is a novel explanation technique that is a special case of RDE. CartoonX first executes RDE in the discrete wavelet position-scale domain of an image $x$ and then visualizes the wavelet mask $s_m$ as a pixel-wise smooth picture. Wavelets efficiently represent 2D piece-wise smooth pictures, commonly known as cartoon-like images, along with providing optimum representations for piece-wise smooth 1D functions \cite{RDE}. Algorithm~\ref{CX} illustrates the steps for obtaining CartoonX explanations.
\begin{algorithm}
	\caption{Algorithm for CartoonX} \label{CX}
	\begin{algorithmic}
		\STATE \textbf{Data: Cervix-AID-Net ($\Psi$), image $x \in [0,1]^n$ with channels $c$, and pixels $k$}.
		\STATE \textbf{Parameters:} Number of steps $N$, sparsity level ($\lambda > 0$), distortion $d$, number of noise samples $L$.
		\STATE \textbf{Initialization:} $s_m := [1,...,1]^T$ on $DWT$ coefficients $h = [h_1,..,h_k]^T$ with $x = f(h)$, where $f$ is $DWT^{-1}$.  
		\FOR {$i \leftarrow 1$ \textbf{to} $N$ \textbf{do}}
		\STATE \textbf{Sampling:} Sample $L$ adaptive Gaussian noise samples $k^{(1)},..,k^{(L)} \sim \mathcal{N} (\mu,\sigma^2)$;
		\STATE \textbf{Obfuscations:} Evaluate obfuscations $y^{(1)},..,y^{(L)}$ with $y^{(i)} := f(h\odot s_m+(1-s_m)\odot k^{(i)}$;
		\STATE \textbf{Clipping:} Clip  obfuscations into $[0,1]^n$;
		\STATE \textbf{Distortion:} Approx. estimated distortion $\hat{E}_D(x,s_m,\Psi):= \sum_{l=1}^{L}d(\Psi(x),\Psi(y^{(i)}))^2/L$;
		\STATE \textbf{Loss:} Compute loss, $l(s_m) := \hat{E}_D(x,s_m,\Psi) + \lambda  ||{s_m}||_1 $;
		\STATE \textbf{Updation:} Update mask $s_m$ with gradient descent step using $\Delta_{s_m}l(s_m)$ and clip $s_m$ back to $[0,1]^k$;
		\ENDFOR
		\STATE \textbf{$\textbf{DWT}$ coefficients:} Get $DWT$ coefficients $\hat{h}$ for greyscale image $\hat{x}$ of $x$;
		\STATE \textbf{Set:} $\xi := f(\hat{h} \odot s_m)$;
		\STATE \textbf{Clipping:} Clip $\xi$ to $[0,1]^k$.
	\end{algorithmic}
\end{algorithm} 
\section{Experimental setup and Results}\label{sec4}
This section covers experimental setup, including evaluation criteria and
evaluation metrics and results. 
\subsection{Experimental setup}
The experiment was performed on an AMD Ryzen 7 PRO 6850U with Radeon Graphics with a frequency of 2.70 GHz. The experimental setup used a 64-bit operating system, x64-based processor, and 16 GB RAM. We used Jupyter Notebook and Python version 3.11.4 to build the model. The high-risk class contains 1404 images, while the low-risk class contains 1749 images. We reshaped all the images to 224 x 224, as the models processes an input image of 224 x 224. 
We used an ``Adam" optimizer,  loss function as ``sparse categorical cross entropy", and accuracy as an evaluation matrix to compile the model. The model iterates for 25 epochs with a batch size of 32. 

\subsection{Results}
The proposed Cervix-AID-Net model is trained and evaluated in multiple scenarios for comprehensive data analysis. To have a faithful comparison concerning performance and tuning parameters, we compared the performance of the proposed model with  benchmark CNN models like AlexNet, GoogleNet (G-Net), and ECANet, respectively  \cite{AlexNet, gNet, ECANet}. AlexNet is a simple eight-layered CNN model capable of performing 1000 class classifications, G-Net includes inception layers, and ECANet is an efficient channel attention module for the CNN model. Table~\ref{tab1} shows the accuracy of Cervix-AID-Net as compared with the benchmark models, using HO and 10-FCV techniques. The results show that our proposed Cervix-AID-Net surpasses the performance obtained using benchmark CNN models for both, HO and 10-FCV validation techniques.  
\begin{table}  \linespread{1.2}\selectfont\centering
	\centering
	\caption{Accuracy comparison of the proposed Cervix-AID-Net with benchmark CNN models.}  \label{tab1}   
	\begin{tabular}   { c | c |c}
		\hline\hline 	
		Models	&	HO	&	10-FCV	\\ \hline  \hline
		AlexNet	&	95.45	&	98.07	\\ \hline
		G-Net	&	97.98	&	98.13	\\ \hline
		ECANet	&	98.82	&	98.73	\\ \hline
		\textbf{Cervix-AID-Net}	&	\textbf{99.33}	&	\textbf{99.81}	\\ \hline \hline
		
	\end{tabular} 
\end{table}
To further evaluate, validate, and compare the performance of the Cervix-AID-Net model, we obtained evaluation matrices and compared them with the performance of benchmark CNN models. Table~\ref{tab2} indicates the performance report of various models using the HO validation technique. It is evident from Table~\ref{tab2} that the proposed Cervix-AID-Net model surpasses most other models in terms of accuracy, sensitivity, F1 measure, MCC, NPV, and balanced accuracy. However, ECANet and G-Net provide slightly better precision, specificity, and FPR than the Cervix-AID-Net model. However, the main limitation of the HO validation technique is that model performance depends significantly on the random split. Also, training uses only a portion of the data, which increases the probability of over-fitting and bias. Therefore, to avoid instability of sampling (where different results are obtained when the experiment is repeated with a new division), we validated our model performance using the 10-FCV technique. Table~\ref{tab2} provides the performance of the benchmark CNN models and the proposed Cervix-AID-Net model using the 10-FCV technique. Table~\ref{tab2} confirms that our developed Cervix-AID-Net model is superior to other benchmark techniques. The Cervix-AID-Net model yielded the highest performance of all the evaluation metrics. Therefore, analysis shows that our Cervix-AID-Net model is robust and accurate over existing CNN models. 
\begin{table}  \linespread{1.2}\selectfont\centering
	\centering \setlength\tabcolsep{2.5pt} 
	\caption{Performance of the proposed Cervix-AID-Net obtained using HO and 10-FCV validation techniques.}  \label{tab2}   
	\tiny
	\begin{tabular}   { c || c |c| c |p{0.7cm} || c |c| c |p{0.7cm}}
		\hline\hline 	
		& \multicolumn{4}{c}{HO}  &\multicolumn{4}{||c}{10-FCV} \\ 	\hline\hline 	
		Models	&	AlexNet	&	G-Net	&	ECANet	&	Cervix-AID-Net	&	AlexNet	&	G-Net	&	ECANet	&	Cervix-AID-Net	\\ \hline \hline 
		ACC	&	95.45	&	97.98	&	98.82	&	\textbf{99.33}	&	98.07	&	98.13	&	98.73	&	\textbf{99.81}	\\ \hline
		SEN	&	99.16	&	100	&	98.47	&	\textbf{98.85}	&	97.06	&	98.63	&	99.21	&	\textbf{99.86}	\\ \hline
		SPF	&	92.98	&	96.52	&	99.10	&	\textbf{99.70}	&	98.90	&	97.74	&	98.36	&	\textbf{99.77}	\\ \hline
		PRC	&	90.38	&	95.38	&	98.85	&	\textbf{99.62}	&	98.65	&	97.15	&	97.93	&	\textbf{99.72}	\\ \hline
		F1	&	94.57	&	97.64	&	98.66	&	\textbf{99.23}	&	97.85	&	97.88	&	98.57	&	\textbf{99.79}	\\ \hline
		MCC	&	0.91	&	0.96	&	0.98	&	\textbf{0.99}	&	0.96	&	0.96	&	0.97	&	\textbf{1}	\\ \hline
		FPR	&	7.02	&	3.48	&	0.90	&	\textbf{0.30}	&	1.10	&	2.26	&	1.64	&	\textbf{0.23}	\\ \hline
		NPV	&	99.40	&	100	&	98.80	&	\textbf{99.10}	&	97.60	&	98.91	&	99.37	&	\textbf{99.89}	\\ \hline
		B-ACC	&	96.07	&	98.26	&	98.78	&	\textbf{99.28}	&	97.98	&	98.18	&	98.78	&	\textbf{99.81}	\\ \hline	 \hline

	\end{tabular} 
\end{table}

The effectiveness of the proposed Cervix-AID-Net model over existing benchmark CNN models was compared in terms of network architecture. Table~\ref{tab5} represents a comparison of tuning parameters. It is evident from Table~\ref{tab5} that our coined model has the least tuning parameters compared to the existing AlexNet, G-Net, and ECANet models. The  AlexNet, G-Net, and ECANet models require 98.11 MB, 28.31 MB, and 58.99 MB of space, respectively. The analysis also reveals that the Cervix-AID-Net model required only 18.04 MB of storage, the least of all the models used for the comparison. Therefore, our proposed model is lightweight and simple. 
\begin{table}  \linespread{1.2}\selectfont\centering
	\centering
	\caption{Comparison of tuning parameters of the proposed Cervix-AID-Net with benchmark CNN models.}  \label{tab5}   
	\begin{tabular}   { c | c |c| c | c }
		\hline\hline 	
		Models	&	TTP	&	TNP	&	NTNP	&	Size (MB)	\\ \hline \hline
		AlexNet	&	25719386	&	25699786	&	19600	&	98.11	\\ 
		G-Net	&	7420982	&	7420470	&	512	&	28.31	\\ 
		ECANet	&	15464090	&	15463130	&	960	&	58.99	\\ 
		\textbf{Cervix-AID-Net}	&	\textbf{4729208}	&	\textbf{4727480}	&	\textbf{1728}	&	\textbf{18.04}	\\ \hline \hline

	\end{tabular} 
	
	\vspace{1mm}
	\tiny{TTP: Total parameters; 	TNP: Trainable parameters;	NTNP: Non trainable parameters}
\end{table}

\begin{figure}
	\centering
	\begin{subfigure}[t]{0.8\textwidth}
		\centering
		\includegraphics[width=0.7\textwidth]{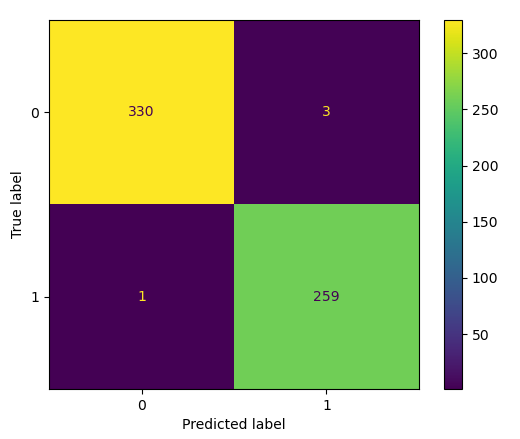}
		\caption{HO validation}
	\end{subfigure}%
	
	\begin{subfigure}[t]{0.8\textwidth}
		\centering
		\includegraphics[width=0.7\textwidth]{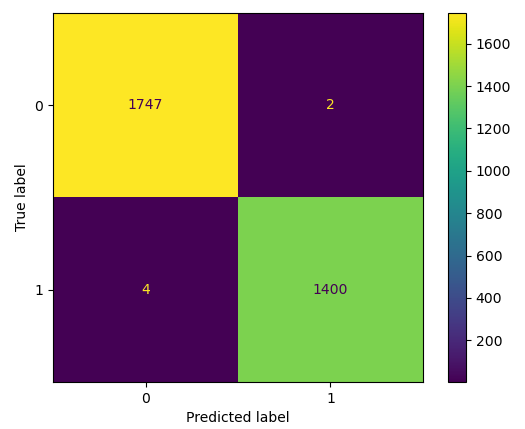}
		\caption{10-FCV}
	\end{subfigure}
	\caption{Confusion matrix obtained for the proposed Cervix-AID-Net model (0-Low-risk and 1-High-risk).}
	\label{cm}
\end{figure}

We also evaluated the confusion matrix to get insight into the proposed Cervix-AID-Net model. Figure~\ref{cm} shows the confusion matrix of the Cervix-AID-Net obtained using HO and 10-FCV techniques. As indicated in Figure~\ref{cm} (a), one image of low-risk has been classified as high-risk, while six images of high-risk class have been identified as low-risk in HO validation. Similarly, for the 10-FCV technique, our developed Cervix-AID-Net model has correctly identified 1747 and 1400 images corresponding to low-risk and high-risk, respectively. However, out of 1749 low-risk instances, two images were incorrectly identified as high-risk, and from 1404 instances of high-risk, four images were incorrectly identified as low-risk. This indicates that our developed model provides an effective classification of targeted classes.  

\begin{figure}[!htbp]
	\centering
	\subfloat[\centering HO validation]{{\includegraphics[width=0.5\textwidth]{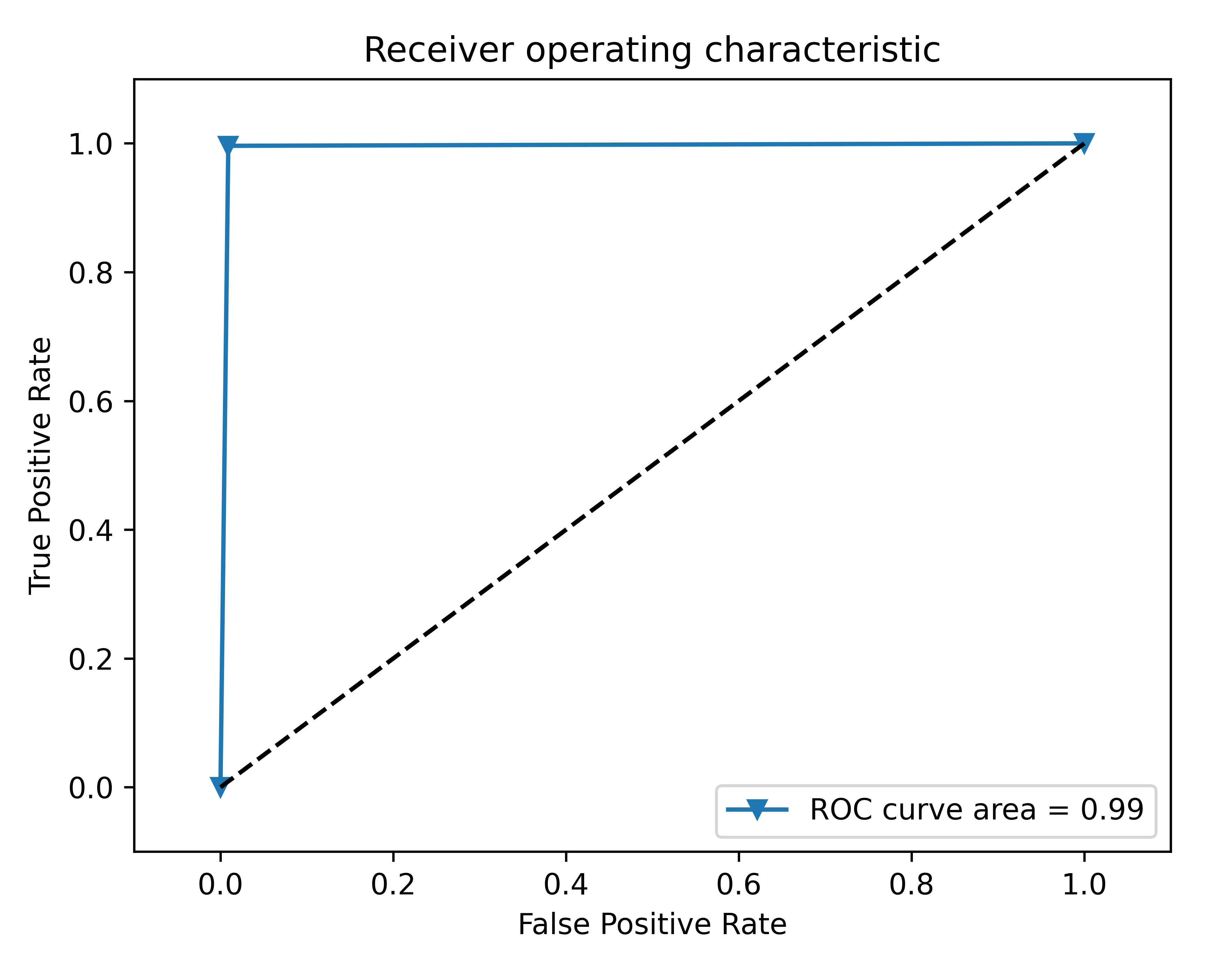}}}
	\qquad
	\centering
	\subfloat[\centering 10-FCV]{{\includegraphics[width=0.5\textwidth]{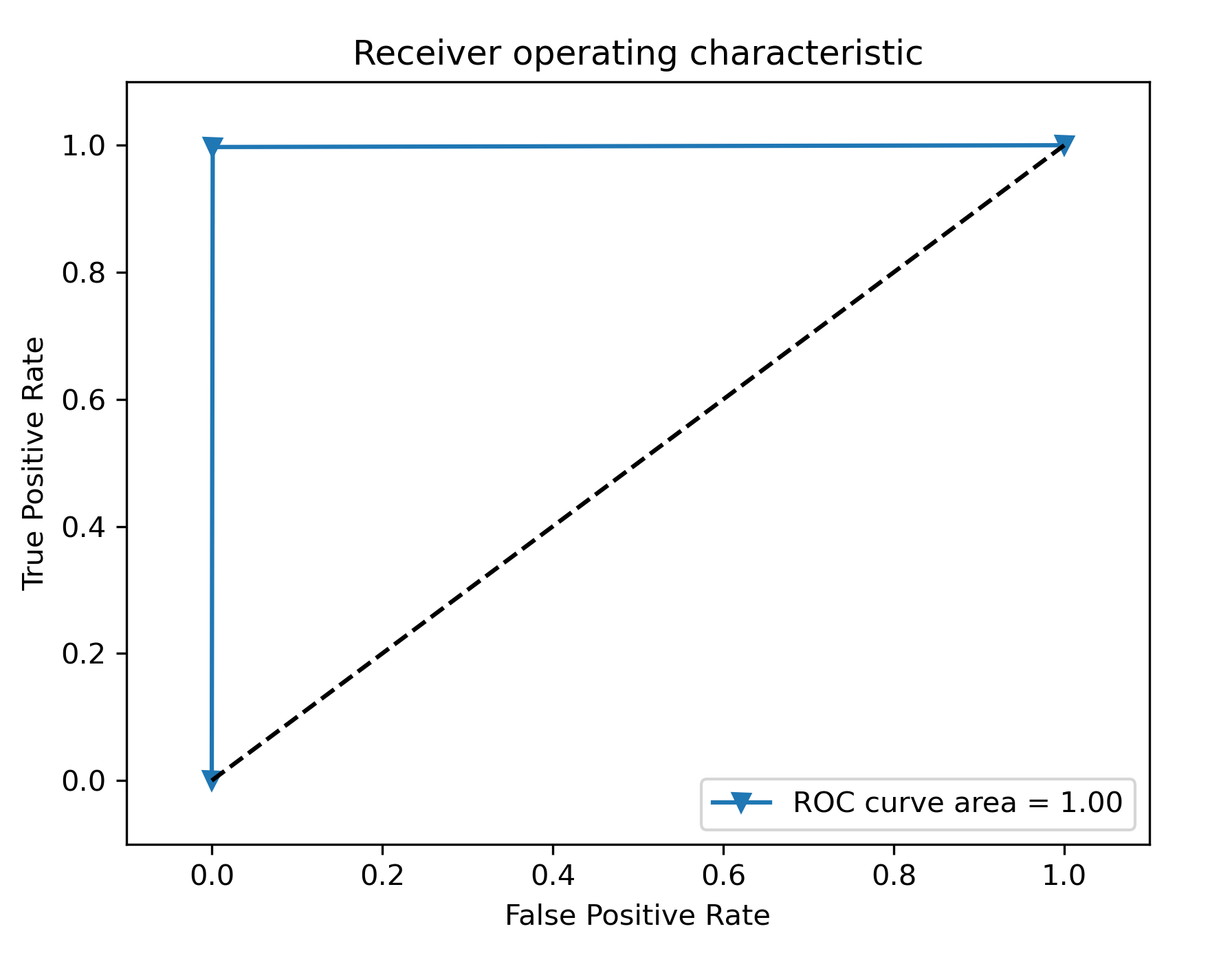}}}
	\caption{ROC-AUC obtained for the Cervix-AID-Net model.}
	\label{roc}
\end{figure}

We further evaluated receiver operator characteristics (ROC) and the area under the curve (AUC) to investigate the binary classification prediction performance of our proposed Cervix-AID-Net model. Figure~\ref{roc} indicates the ROC and the AUC curve obtained using HO and 10-FCV techniques. Our proposed model has obtained 99\% and 100\% of AUC for HO and 10-FCV, which confirms the binary classification ability of the Cervix-AID-Net model.  

\section{Discussion}\label{sec5}
\begin{figure}[!htbp] 
	\centering
		\includegraphics[width=0.8\textwidth]{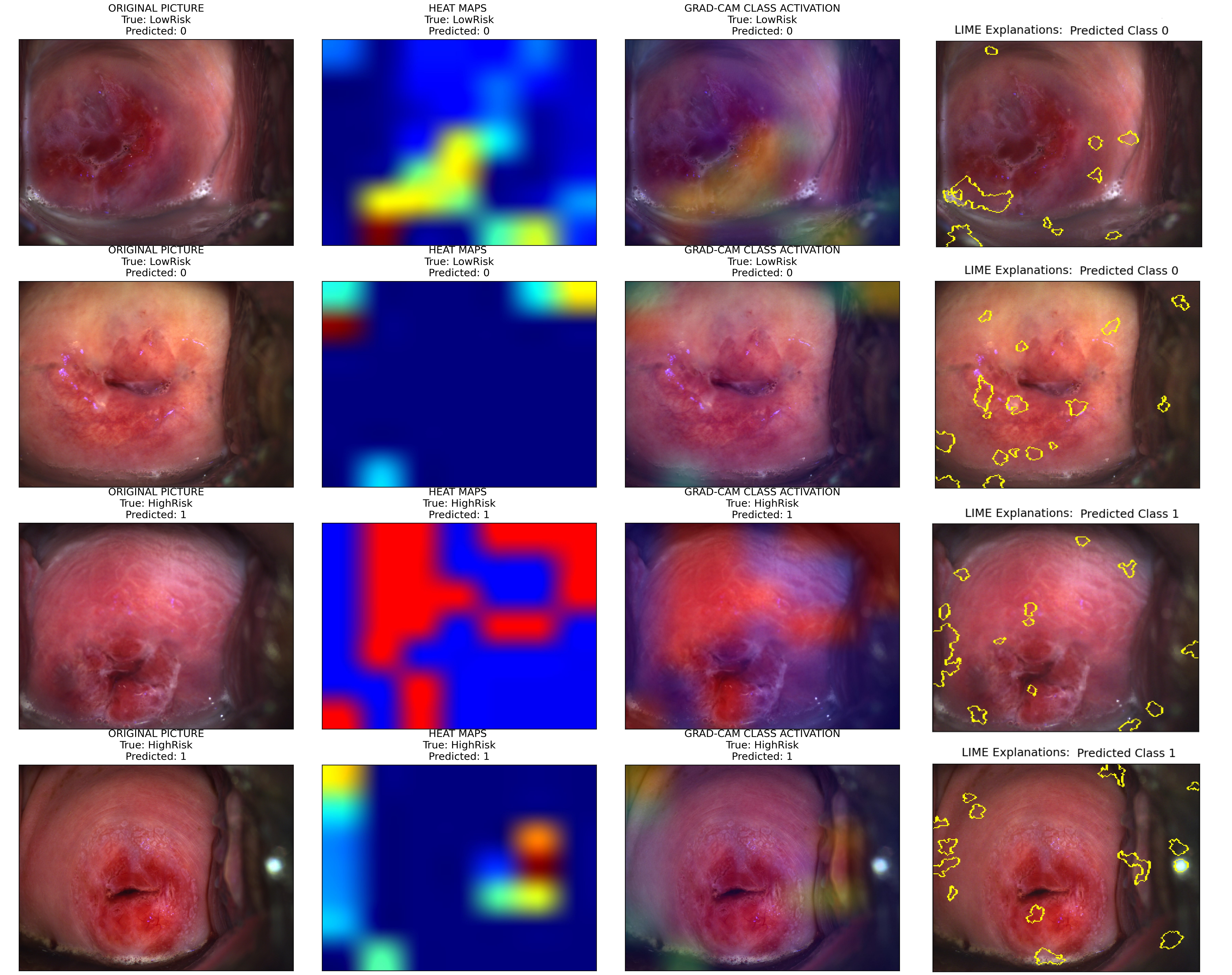}
	\caption{Example of grad-CAM and LIME explanations obtained from the last convolutional layer of the proposed Cervix-AID-Net model. (0-Low-risk and 1-High-risk)}
	\label{gcam}
\end{figure}

\begin{figure}[!htbp]
	\centering
	\includegraphics[width=0.8\textwidth]{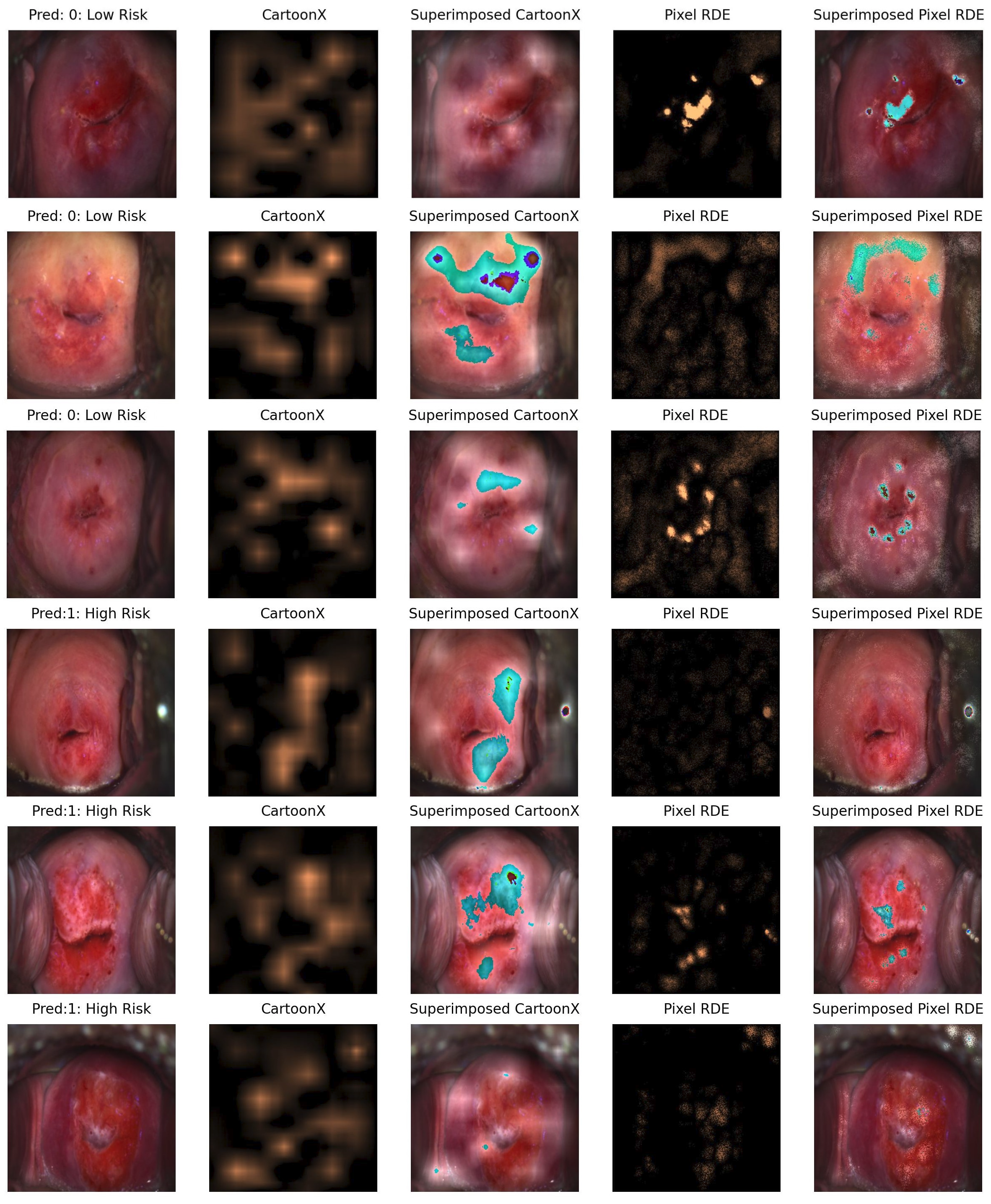}
	\caption{Example of explanations obtained from CartoonX and pixel RDE}
	\label{rde}
\end{figure}

Our Cervix-AID-Net model aims to provide transparent and effective explanations for its decisions using four explainable techniques. These XAI techniques require tuning of hyper-parameters to generate explanations for the model's decisions. Table~\ref{xai} represents various tuning parameters used for LIME, pixel RDE, and CartoonX.
\begin{table}  \linespread{1}\selectfont\centering
	\centering
	\caption{Tuning parameters used for XAI methods.}  \label{xai}   
	\begin{tabular}   { c | c |c| c }
		\hline\hline 	
		Methods	&	LIME	&	Pixel RDE	&	CartoonX	\\ \hline\hline
		Number of features	&	20	&	--	&	--	\\ \hline
		Number of samples	&	2000	&	--	&	--	\\ \hline
		Kernel size	&	1	&	--	&	--	\\ \hline
		Max distance	&	200	&	--	&	--	\\ \hline
		Ratio	&	0.2	&	--	&	--	\\ \hline
		Segmenter	&	Quick shift	&	--	&	--	\\ \hline
		Lambda	&	--	&	4	&	285	\\ \hline
		Step size	&	--	&	0.01	&	0.1	\\ \hline
		Number of steps	&	--	&	200	&	100	\\ \hline
		Batch size	&	--	&	--	&	16	\\ \hline
		Obfuscation 	&	--	&	Gaussian	&	Gaussian	\\ \hline\hline
		
	\end{tabular} 
	
	\vspace{1mm}
	
\end{table}
Grad-CAM provides explanations based on output feature maps that generate the gradients from the last convolutional layer. Fig.~\ref{gcam} indicates the grad-CAM obtained from the last convolutional layer of our proposed Cervix-AID-NET model. Our analysis confirms that the heat maps from grad-CAM outline the relevant regions around the cervix. However, there are some instances where the network does not outline any regions around the cervix. It could be because of uneven light distribution on the images, motion artifacts, or instances of outliers, shown in the second image in Fig.~\ref{gcam}. On the other hand, LIME explains specific instances using input features. LIME approximates the pre-trained model to generate a surrogate model that explains local approximations around the input image. As evident from Fig.~\ref{gcam}, LIME highlights crucial regions in the neighborhood of the input image, which provides the most relevant input features for that particular instance. Fig.~\ref{gcam} reveals that LIME maps the region around the cervix, marking relevant features to classify correctly in high-risk or low-risk class, which are significant for that instance. However, LIME fails to identify all the features in the image which are relevant for decision-making. Our observation with LIME was that though it does not highlight all the regions accurately, LIME does not miss out on marking the crucial features around the cervix. LIME's explanation focuses primarily on the characteristics that contribute independently to the model output, and their contributions should reflect their importance. However, such explanations are relevant and crucial in scenarios like segmentation, where individual pixels carry relevant and interpretable meanings for the decision. However, for a classification problem, it is desired to identify a group of features that are interpretable but not individual pixels that do not carry any interpretable meaning. Therefore, we favour pixel RDE and CartoonX approaches, which seek a group of significant characteristics rather than an estimate of individual relative contributions. Fig~\ref{rde} provides an illustrative representation of pixel RDE and CartoonX. The explanation reveals that the mask provided by pixel RDE is the explanation as it lies in pixel space. Our analysis indicates that explanations of pixel RDE are highly non-stationary. At some instances, pixel RDE focuses on the relevant regions as indicated by clinical experts, as highlighted in the images of the first, third, fifth, and sixth rows. However, in some instances, the explanations yielded by pixel RDE were nowhere near the cervix region because of light intensities or outliers, as indicated in the images of the second and fourth rows. Finally, Fig~\ref{rde} shows that explanations yielded by CartoonX methods are the most relevant as they map the crucial region in the image. Our analysis shows that explanations given by CartoonX lie around the cervix region, which is the most crucial part of the image. The reason for a more meticulous explanation of CartoonX is its ability to extract relevant piece-wise smooth parts of an image instead of relevant pixel sparse regions. The sparsity in the wavelet domain captures interpretable input features from the image compared to sparsity in the pixel domain, instance-based explanations, and output neuron activations. Our analysis shows that CartoonX captures piece-wise smooth explanations that can reveal relevant piece-wise smooth patterns that are not easily visible with existing grad-CAM that operates on the output feature maps and pixel-based methods like LIME and pixel RDE.   

Noise and blur are two prominent image classification issues that might impair the performance of ML/DL models. Understanding and addressing these challenges is critical for enhancing image classifier robustness and accuracy. Random noise defines the existence of random fluctuations in pixels that do not constitute part of the image's actual content. It may be introduced during acquisition, transmission, or processing and can influence image quality and appearance. Blur occurs when an image's features are not finely defined, commonly caused by motion during acquisition, defocus, or other optical difficulties. To test these challenges, we added different levels of Gaussian noise to the test images. We randomly selected eight images and added 3\%, 5\%, 10\%, and 30\% of Gaussian noise. We also added a median blur of 5\%, 10\%, 20\%, and 30\% to these images. Analysis shows that the model performance remains unchanged with Gaussian noise of 3\%, and blur of 5\%, and 10\%. However, the accuracy drops by 25\% with 5\% of Gaussian noise and 20\% of blur. In addition, the performance of the developed model reduces by 50\% for 10\% and 30\% of Gaussian noise and 20\% blur. The results indicate that the performance of our developed model does not change until a random noise exceeds 3\% and until a blur of 10\%. The analysis shows that the proposed model can work with low-resolution images with minimal loss.     

\begin{figure}[!htbp]
	\centering
	\centering
	\subfloat[\centering Gaussian Noise]{{\includegraphics[width=0.8\textwidth]{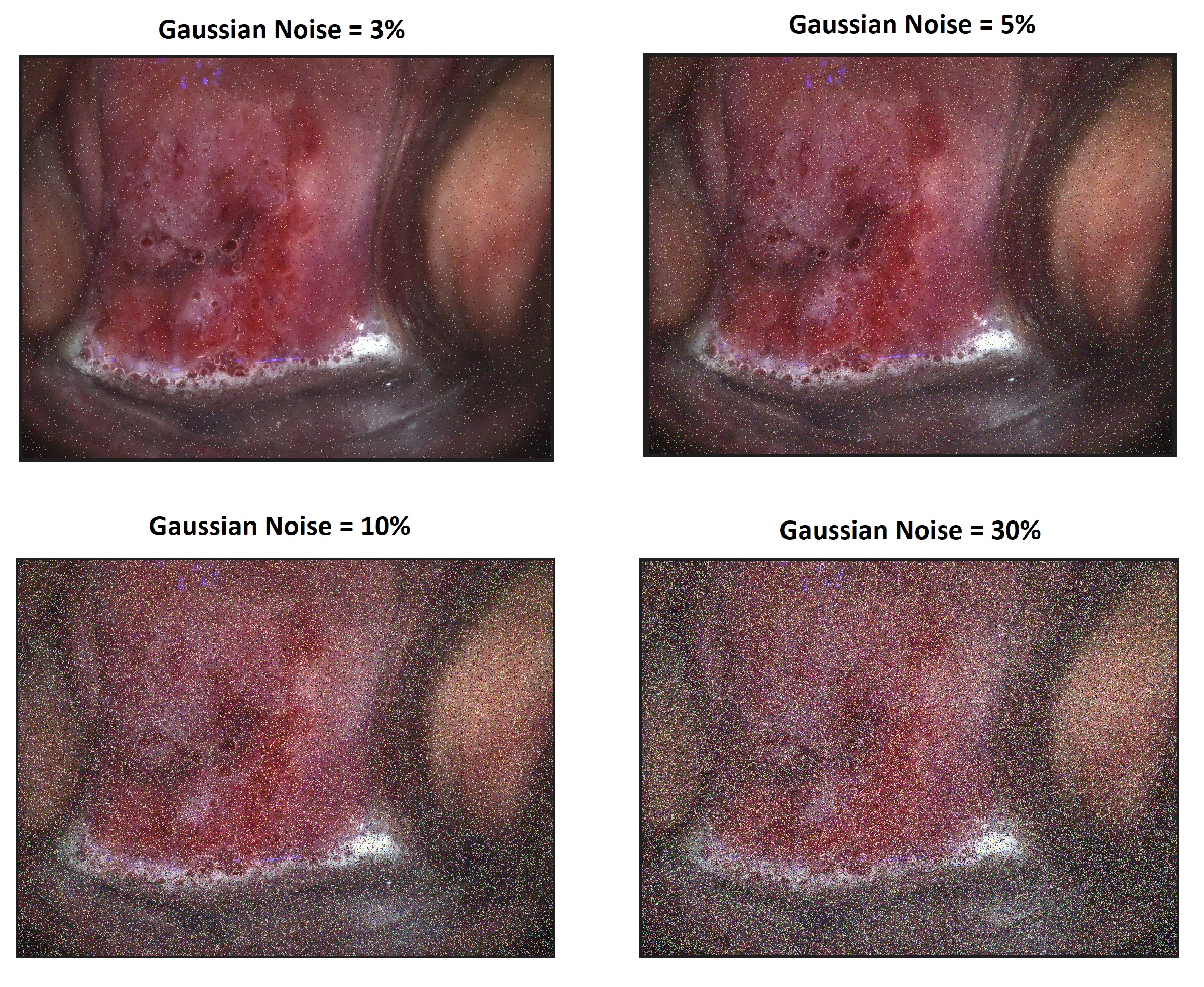}}}
	\qquad	
	\centering
	\subfloat[\centering Median blur noise]{{\includegraphics[width=0.8\textwidth]{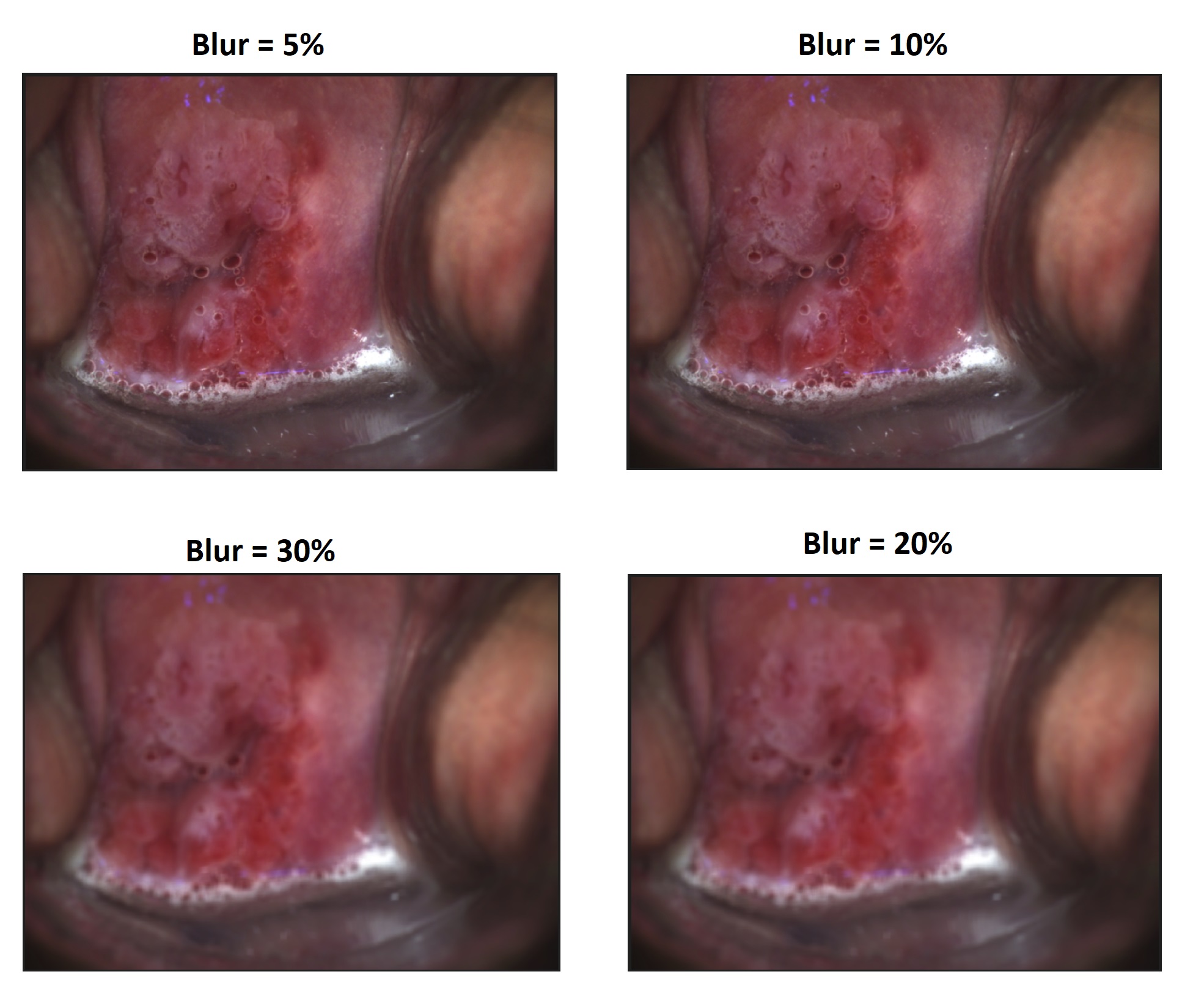}}}
	\caption{Example of different noise sources with various levels of noise disturbances.}
	\label{noise}
\end{figure}
\begin{table*}  \linespread{1.3}\selectfont\centering
	\centering \setlength\tabcolsep{2.5pt} 
	\tiny
	\caption{Performance comparison of the Cervix-AID-Net with existing SOTA techniques employing binary classification.}  \label{tab6}   
	\begin{tabular}   { c | c |c| c | c| c| c| c| c| c| c }
		\hline\hline 	
		Author \& Year	&	Images	&	Image-generating devices	&	Classes	&	Method	&	Validation	&	ACC	&	PRC	&	Recall	&	SPF	&	NPV	\\ \hline \hline
		Li et al. \cite{1} (2020)	&	7668	&	Traditional colposcopy	&	Pos vs Neg	&	E-GCN	&	HO (80:20)	&	81.95	&	81.97	&	81.78	&	--	&	--	\\ \hline
		Elakkiya et al. \cite{2} (2022)	&	3105	&	Traditional colposcopy	&	Normal vs AN	&	FR-CNN-GAN	&	HO (80:20)	&	98.55	&	98.66	&	100	&	100	&	--	\\ \hline
		Adweb et al. \cite{3} (2021)	&	7920	&	Traditional colposcopy	&	HC vs PC	&	ResNet 	&	HO (60:40)	&	100	&	100	&	100	&	100	&		\\ \hline
		Kim et al. \cite{5} (2022)	&	--	&	Digital colposcopy	&	--	&	AIDOTNet 	&	--	&	--	&	81.13	&	74.14	&	83.05	&	--	\\ \hline
		Saini et al. \cite{6} (2020)	&	800	&	Traditional colposcopy	&	Type 1 vs Type 2	&	ColpoNet 	&	HO (70:15:15)	&	81.35	&	--	&	--	&	--	&	--	\\ \hline
		Kim et al. \cite{11} (2013)	&	2000	&	Cervicography (discontinued)	&	Normal vs  
		CIN2+	&	Texture features and SVM	&	10F-CV	&	--	&	--	&	75	&	76	&	--	\\ \hline
		Cho et al. \cite{12} (2020)	&	791	&	Digital colposcopy	&	HighRisk vs LowRisk	&	Inception-ResNet 	&	HO (85:15)	&	69.3	&	47.2	&	66.7	&	70.6	&	84	\\ \hline
		\multirow{2}{*}{Liu et al. \cite{13} (2021)}	&	\multirow{2}{*}{15276}	&	\multirow{2}{*}{Digital colposcopy}	&	NC vs LSIL

		&	\multirow{2}{*}{ResNet-24}	&	\multirow{2}{*}{HO (70:10:20)}	&	88.2
		
		&	85.3
		
		&	90.1
		&	86.7
		&	91
		\\
		&		&		&	HSIL vs HSIL+	&		&		&	 79.7	&	 60.6	&	80.2	&	79.6	&	91.1	\\ \hline
		Sim{\~o}es et al. \cite{17} (2014)	&	170	&	Digital colposcopy	&	--	&	MLP	&	--	&	72.15	&	--	&	69.78	&	68	&	--	\\ \hline
		Asiedu et al. \cite{18} (2019)	&	--	&	Digital colposcopy	&	Pos vs Neg	&	Gabor segmentation-SVM	&	5F-CV	&	80	&	--	&	81.3	&	78.6	&	--	\\ \hline
		Miyagi et al. \cite{19} (2020)	&	253	&	Traditional colposcopy	&	LSIL vs HSIL	&	CNN 	&	HO (80:20)	&	94.1	&	97.7	&	95.6	&	83.3	&	71.4	\\ \hline
		Song et al. \cite{20} (2015)	&	60000	&	Cervicography (discontinued)	&	Pos vs Neg	&	Multimodal CNN	&	10F-CV	&	89	&	--	&	83.21	&	94.79	&	--	\\ \hline
		Miyagi et al. \cite{21} (2019)	&	310	&	Traditional colposcopy	&	CIN1 vs CIN2+	&	CNN 	&	5F-CV	&	82.3	&	94.7	&	80	&	88.2	&	62.5	\\ \hline \hline
		\multirow{2}{*}{\textbf{Our Cervix-AID-Net}}	&	\multirow{2}{*}{\textbf{3154}}	&	\multirow{2}{*}{\textbf{Digital colposcopy}}	&	\multirow{2}{*}{\textbf{HighRisk vs LowRisk}}	&	\multirow{2}{*}{\textbf{CNN with CBAM}}	&	\textbf{HO (80:20)}	&	\textbf{99.33}	&	\textbf{99.62}	&	\textbf{98.85}	&	\textbf{99.70}	&	\textbf{99.10}	\\ 
		&		&		&		&		&	\textbf{10F-CV}	&	\textbf{99.81}	&	\textbf{99.72}	&	\textbf{99.86}	&	\textbf{99.77}	&	\textbf{99.89}	\\ \hline \hline

	\end{tabular} 
	
\end{table*}
Finally, the effectiveness of the developed Cervix-AID-Net model is test by comparing the performance with existing state-of-the-art (SOTA) techniques. Table~\ref{tab6} represents the binary classification evaluation report of the SOTA techniques, that is, the best-performing model or algorithm that achieves the highest accuracy or provides the most advanced functionality. The comparison report of performance parameters shows that our developed model is superior. Hence, our Cervix-AID-Net model is robust, accurate, and effective in classifying high-risk and low-risk cervical precancer images.  The key features of the proposed model are listed below:
\begin{itemize}
	\item The dataset is unique due to the exact mapping of the cervix.
	\item The Cervix-AID-Net model is accurate and explainable for classifying high-risk and low-risk cervical precancer.
	\item The model is lightweight due to fewer tuning and learnable parameters over benchmark CNN networks.
	\item The Cervix-AID-Net model generates highly discriminant features due to the CBAM module.
	\item The proposed model is robust due to validation using HO and 10-FCV techniques.
	\item The proposed Cervix-AID-Net is accurate as it has obtained 99.81\% accuracy, outperforming SOTA techniques.
\end{itemize}
However, the proposed model has few short-comings, which are mentioned below:
\begin{itemize}
	\item The model was tested on a relatively small dataset from a single hospital.
	\item The effects of domain shift are not verified due to the unavailability of external datasets.
	\item Only binary classification is targeted.
	
\end{itemize}
In the future, we will explore the following research directions to further improve our model, by addressing: 
\begin{itemize}
	\item Segmentation of low-risk and high-risk cervical precancer.
	\item Uncertainty quantification using different noise sources.
	\item Validation of model’s performance on an external dataset.
	\item Hyper-parameter tuning for XAI techniques. 
	
\end{itemize}

\section{Conclusion}\label{sec6}
The paper presents an effective AI framework for the classification of low-risk and high-risk cervical precancer lesions from colposcopy images. The proposed Cervix-AID-Net model captures representation features from the colposcopic images by maintaining attention over the channel and spatial domain. Our proposed model has effectively identified \textit{what} is the significant region in the image and \textit{where}  it is located within the image using CBAM. Therefore, our Cervix-AID-Net model yielded the highest accuracy in classifying high-risk and low-risk cervical precancer instances, compared to existing benchmark CNN models. In addition, our model facilitates visually readable transparent explanations for decision-making, which makes it robust and effective. The proposed model explains relevant output feature maps and important input features required for accurate decision-making. We show that piece-wise smooth explanations extract representative and discriminant features of the input image. Thus, the explanations yielded by CartoonX were the most interpretable and discriminant compared to pixel RDE, LIME, and grad-CAM techniques. The experimental analysis demonstrated that the proposed model may yield effective decision-making on high-resolution and low-resolution images. The Cervix-AID-Net model is
lightweight, has been tested on multiple scenarios were it yielded the best performance matrices compared to current models, and it has integrated XAI techniques. In conclusion, the Cervix-AID-Net model, presented in this article, is an accurate, robust, transparent, effective, and simple, that could be used in future for early identification and assessment of cervical precancer lesion during colposcopy.

\section{Funding Information}
For this study we got funding from the BetaHealth Program from NOVO Nordisk (ID-nummer 2023-1384) and an Innovation Grant from Region Syddanmark.
\bibliographystyle{unsrt}  
\bibliography{mybibn.bib}

\end{document}